\documentstyle[12pt,psfig]{article}
\newcommand{\be}{\begin{equation}}
\newcommand{\ee}{\end{equation}}
\newcommand{\bea}{\begin{array}{c}}
\newcommand{\eaa}{\end{array}}
\newcommand{\ba}{\begin{eqnarray}}
\newcommand{\ea}{\end{eqnarray}}

\begin{document}
\date{}
\title{Canonical Description of Strangeness Enhancement\\ from\\ p-A
to  Pb-Pb Collisions}
\author{ Salah Hamieh$^{\rm (a)}$, Krzysztof Redlich$^{\rm (a,b)}$  
and Ahmed Tounsi$^{\rm (a)}$}
\maketitle
%
{\centerline {$\rm (a)$Laboratoire de Physique Th\'eorique et Hautes Energies,}}
{\centerline{
Universit\'e Paris 7, 2 pl Jussieu, F--75251 Cedex 05}}

{\centerline{
${\rm (b)}$Institute of Theoretical Physics, University of
Wroc\l aw,
}}\centerline{ PL-50204 Wroc\l aw, Poland}

 \vskip 0.7cm

\begin{abstract}
We consider the production of strange particles in  Pb-Pb and p-A collisions
at the SPS energy reported by the WA97 experiment.
We show that the observed enhancement of strange baryon and antibaryon yields 
in Pb-Pb collisions relative to  p-Be and p-Pb can be 
explained in terms of the statistical model
formulated in  canonical ensemble with respect to strangeness conservation.
The importance and the role of strangeness under saturation  is also
discussed. 

\end{abstract}

\section{Introduction}

Various experiments at CERN SPS reported an enhancement of strange
particle production when going from p-p, p-A to A-A collisions \cite{qm99ski,Ant99,Lie99,And298,Sik99}.
In particular the recent analysis of  Pb-Pb reaction 
by the WA97 collaboration has shown, by comparing
 p-Be, p-Pb with Pb-Pb, a strong enhancement in the pertinent
 strange and antistrange baryon yields \cite{Ant99,Lie99,And298}.
 In addition
this enhancement was found to increase with strangeness content
of the particle.
\par
In this letter we show that the observed features
 of the data can be well described
by the thermal model for particle production using 
 canonical formulation
of strangeness conservation.
\par
The statistical models formulated in the grand canonical (GC) ensemble, 
with respect to conservation laws, were shown to give a  satisfactory 
description of particle production in 
heavy ion collisions at SPS and AGS energies 
\cite{Bra99,Bra95,CLK,he,Let95}.
 However, when applying this model to the low energy heavy ion \cite{Cle99}, or high energy 
hadron-hadron or $e^+e^-$ reactions one necessarily needs to treat
the conservation laws exactly \cite{Bec96}. 
The exact conservation of quantum numbers, that  is the canonical (C) approach,
is known to reduce the phase space available for particle production
due to additional constraints appearing through requirements of local
quantum number conservation \cite{Hag71,Raf80,Hag85,Cle91}. Thus, the immediate question is to what
extend the canonical suppression can
 account for the observed effect in the WA97 experiment.
\par
The concept of canonical suppression as an origin of strangeness enhancement
has been already addressed in the literature
\cite{Cle96} and also tested in the context of the SPS data \cite{Sol98}.
Indeed, it was shown \cite{Sol98}
that the increase of $K_S^0/h^-$ and $(\Lambda+K)/\pi$ ratios from
p-p to A-A collisions can
  be possibly explained
by the canonical effect. In view of the new data from the
 WA97 collaboration
we will test further this idea and show that the
 suppression of strange particles phase space
when going from GC to C formulation 
of  strangeness conservation can 
quantitatively explain the reported data.
\par
 In the following  section we present the main
aspects  of the statistical model formulated in canonical ensemble
and then we compare the model
  with the experimental data. 

\section{Canonical Formulation of Strangeness Conservation}

 We consider a gas composed of all particles and resonances with  mass up to
$m\simeq 2.4$GeV. Conservation of 
 baryon number is introduced in GC ensemble, thus it is controlled
by baryochemical potential. Strangeness conservation, however,
is introduced on the canonical
 level. Having in mind the application of
 the model to particle
production in heavy ion collisions  we also
 require   that the overall strangeness
of the system is zero. Using the general procedure for the canonical
strangeness conservation \cite{Hag85,Cle91} the partition function of a gas of particles
with strangeness $0,\pm 1,\pm 2,\pm 3$ can be written as follows:

\begin{equation}
Z_{0}={1\over {2\pi}}
       \int_{-\pi}^{\pi}
                   d\phi~ \exp{(\sum_{s=- 3}^3
S_se^{is\phi})},
\label{eq1}
\end{equation}
where $S_s=V_0\sum_i Z_i$ and the sum is taken over all particles
and resonances carrying strangeness $s$. 
For   particle of mass $m_i$, degeneracy factor $d_i$,  baryon number $B_i$,
 and assuming Boltzmann statistics we have:

\begin{equation}
Z_i={d_i\over {2\pi^2 }}m_i^2TK_2(m_i/T)\exp {(B_i\mu /T)}.
\label{eq2}
\end{equation}
\begin{figure}[tb] 
\vspace*{-1.75cm}
\centerline{\hspace*{.5cm}
\psfig{width=
10cm,figure=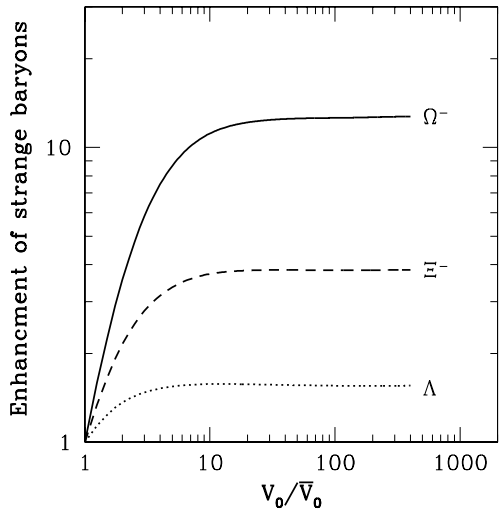}
}
\vspace*{-.4cm}
\caption{ 
Volume dependence of strange baryon densities normalized
to their value calculated in ${V}_0=\bar{V}_0=7.4$ fm$^3$
 for $T=168$ MeV and $\mu=266$ MeV. 
\protect\label{Ape}}
\vspace*{-.2cm}
\end{figure}
With the particular
 form of the partition 
function given by Eq.\,(\ref{eq1})  the  density $n_s$ of particle $i$
with
 strangeness $s$  in volume $V$
 is obtained by the replacement  $Z_i\mapsto \lambda_iZ_i$
in Eq.\,(\ref{eq1})  and then   taking an  appropriate derivative \cite{Hag85,Cle91}:

\begin{eqnarray}
n_{\pm s}&=&{{\langle N_{\pm s}\rangle}\over V}  \equiv{{\lambda_i}\over {V_0}}{{\partial {\ln Z_{0}}}\over
 {\partial {\lambda_i}}}|_{\lambda_i=1}\simeq
Z_{\pm s} {{(S_{\pm1})^s}\over 
{{(S_{+1}}S_ {-1})^{s/2}}}
 \nonumber     \\
      &&\{I_s(x_1)I_0(x_2)+\sum_{m=1}^\infty 
I_m(x_2)[I_{2m+|s|}(x_1)A^{m/2}+I_{2m-|s|}(x_1)A^{-m/2}]\}/Z_0
\label{eq3}
\end{eqnarray}
where
\begin{equation}
x_k\equiv 2\sqrt {S_kS_{-k}}~~,~~
 A\equiv {{S_{-1}^2S_2}\over {S_1^2S_{-2}}},
\end{equation}
and the partition function
\begin{equation}
Z_0\simeq I_0(x_1)I_0(x_2)+
\sum_{m=1}^\infty I_{2m}(x_1)I_m(x_2)[A^{m/2}+A^{-m/2}].
\end{equation}

In the derivation of Eq.\,(\ref{eq3}) we have neglected, after differentiation
over  particle fugacity,
the term $S_{\pm 3}$. This approximation, however, due to 
 small value of 
$S_{\pm 3}$ coefficients in comparison with $S_{\pm 1} ~{\rm and}~ S_{\pm 2}$
 is quite satisfactory. The complete expression for the partition function
without any approximation can be found in reference \cite{Cle91}.

  Due to the locality of  strangeness conservation, 
we have introduced  
two volume parameters, $V$ and $V_0$ in Eq.\,(\ref{eq3}) \cite{Hag71}. The parameter  $V$ describes
the volume of the system at freeze out, thus $V$  is also
 the  overall normalization 
factor
which fixes the particle multiplicity from the corresponding density. 
The volume  $V_0$ is interpreted
as a correlation volume where particles are created to fulfill
the  requirement  of 
local conservation of quantum numbers. It
 was shown \cite{Cle99} that in the 
 canonical description  of particle production in  
low energy heavy ion collisions
 $V\neq V_0$ is required
 to reproduce both the particle multiplicity
 ratios and the  particle yields.
The  volume parameter $V$
 scales with the total number of participants, $V\sim A_{part}$, whereas
$V_0=4\pi R_0^3/3$ where  $R_0\simeq (1.15\pm 0.05) A_p^{1/3}$ with
 $A_p$ being  the  number
of the projectile participants.

Comparing the result of Eq.\,(\ref{eq3}) with the standard expression for the particle
density in  GC ensemble, one can see that the term multiplying 
$Z_{\pm s}$ in  Eq.\,(\ref{eq3}) plays  the role of the strangeness fugacity\ 
\cite{Cle99,Hag85}.
Indeed
taking the limit for large  $x_i$,
  in Eq.\,(\ref{eq3}),
that is  the limit of high temperature and/or  large $V_0$,
 the GC result for the particle density is recovered.
 In the opposite limit the C ensemble leads to a strong
suppression of strange particle densities.
 
In Fig.\ref{Ape} we show the dependence of the density of  strange
particles on $V_0$  for fixed $T=168$ MeV and $\mu=266$ MeV.
In order to illustrate the enhancement with increasing volume and strangeness content of the particle we show in Fig.\ref{Ape}
the density of $\Lambda , \Xi^-$ and $\Omega $ baryons
normalized to its value calculated in the volume $V_0=\bar { V_0}=7.4$ fm$^{3}$,
as a function of $V_0/\bar {V_0}$. It is clear from Fig.\ref{Ape} that
the  particle densities are 
increasing with the size of the system.  For sufficiently large $V_0$
the densities are saturating which 
 indicates that the GC limit is  reached. 
The enhancement is seen to increase with the strangeness content of
the particle and obviously depends on the choice   of $\bar {V_0}$.
Keeping the value of  baryon chemical potential and/or temperature as 
 being $V_0$ independent, 
the enhancement of antibaryons is  similar to the one seen in Fig.\ref{Ape}
for baryons.
Thus, the possible asymmetry in enhancement for strange baryons and 
antibaryons when going from small to large system requires
 $\mu$ and/or $T$ to be $V_0$  dependent.  

In the following section we will discuss to what extend the above model can
be used to understand strangeness enhancement
versus centrality for baryon and antibaryons reported by the
 WA97 collaboration.

\section{Comparison with the Experimental Data}

 It is  seen in Fig.\ref{Ape} that  the model reproduce the main
features  of centrality dependence of strangeness enhancement in the WA97 data.
The quantitative comparison of the model with the experimental data
require, however, a particular   choice of the parameters, that is:		
the temperature $T$, the volume   $V_0$ and the 
baryon chemical potential $\mu$. 
 We  have used for Pb-Pb collisions: $T=168$ MeV and 
$\mu =266$ MeV, the values which have been shown \cite{Bra99} to give
a good description of all data on  particle
 multiplicity ratios measured by the  NA49 and WA97 experiments.
In general  temperature and  chemical potential could vary with 
$V_0$ that is also with  $A_p$. From previous 
analysis we know \cite{clT} that for the same 
collision energy
the temperature variations with   $A_{part}$ is small.
 Thus, we keep
$T$ to be the same for both, p-Be and  Pb-Pb systems.
 The chemical potential, however, 
as discussed in the previous section, has to be taken 
as $A_{part}$ dependent since  
 the data  are showing larger  enhancement for strange  than for 
antistrange baryons. This  indicates  lower value of  $\mu $ in p-Be than
 in Pb-Pb system. We use $\bar\mu$ in p-Be as a free parameter which is chosen
to reproduce the enhancement asymmetry between  $\Xi^-$ and $\bar\Xi^+$ or
simply the $\Xi^- /\bar\Xi^+$ ratio measured in  p-Be collisions.

 Following the interpretation
of $V_0$, described in the previous section, we chose $R_0\sim 1.2$ fm
in p-Be
that is $V_0=\bar{V}_0$. 
To make a quantitative comparison of the model with the experimental data
we are left with only one parameter, the effective chemical potential $\bar\mu$ in p-Be 
which was taken to be 150 MeV. This value is required in comparison
with $\mu =266$ MeV in Pb-Pb in order to obtain the correct splitting 
between baryons and antibaryons of the enhancement factors without
affecting the average enhancement factors.

\begin{figure}[tb]
\vspace*{-1.75cm}
\centerline{\hspace*{-5.6cm}
\psfig
{width=
11cm,figure=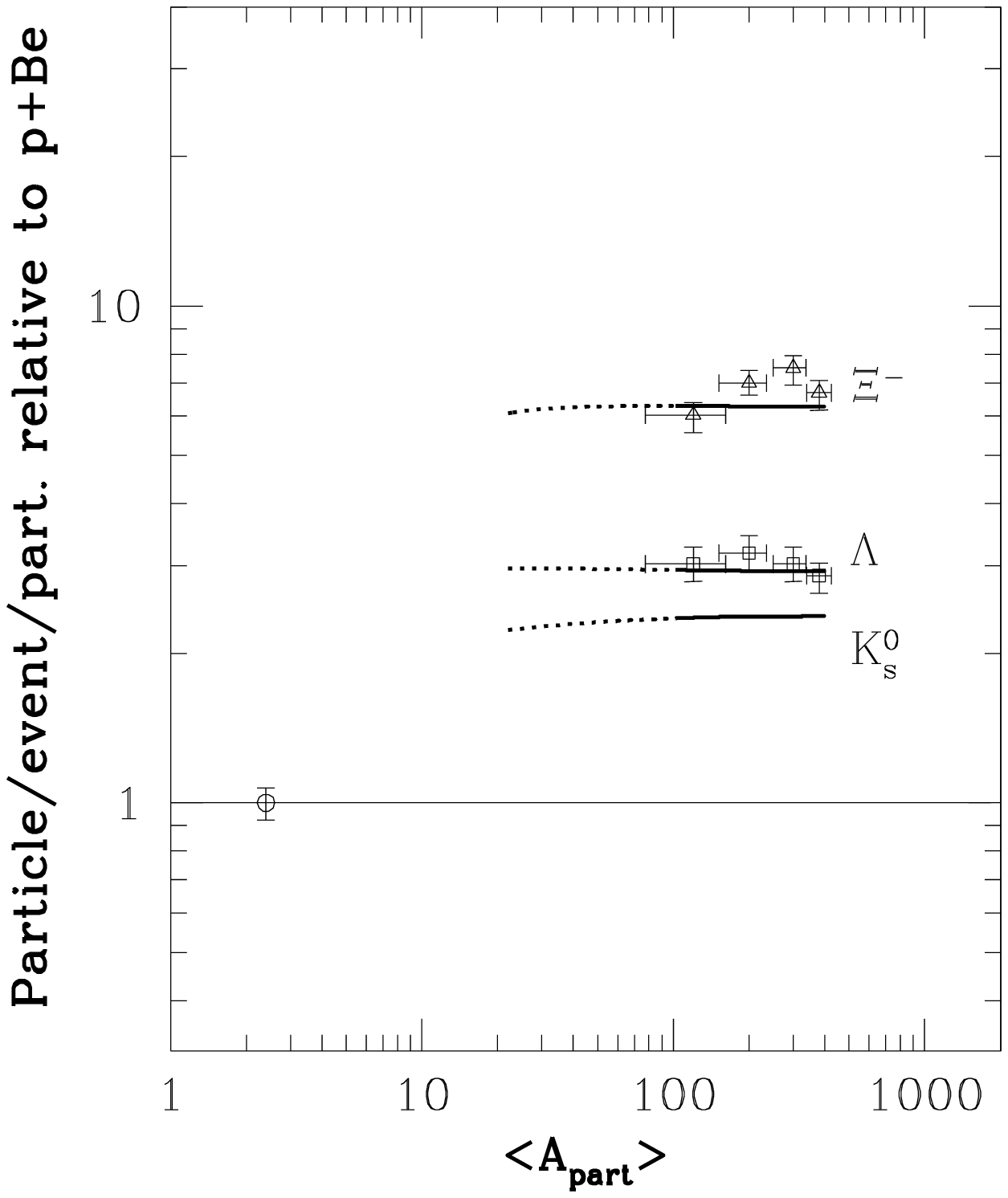}
}
\vspace*{-11.cm}
\centerline{\hspace*{10cm}
\psfig
{width=
11cm,figure=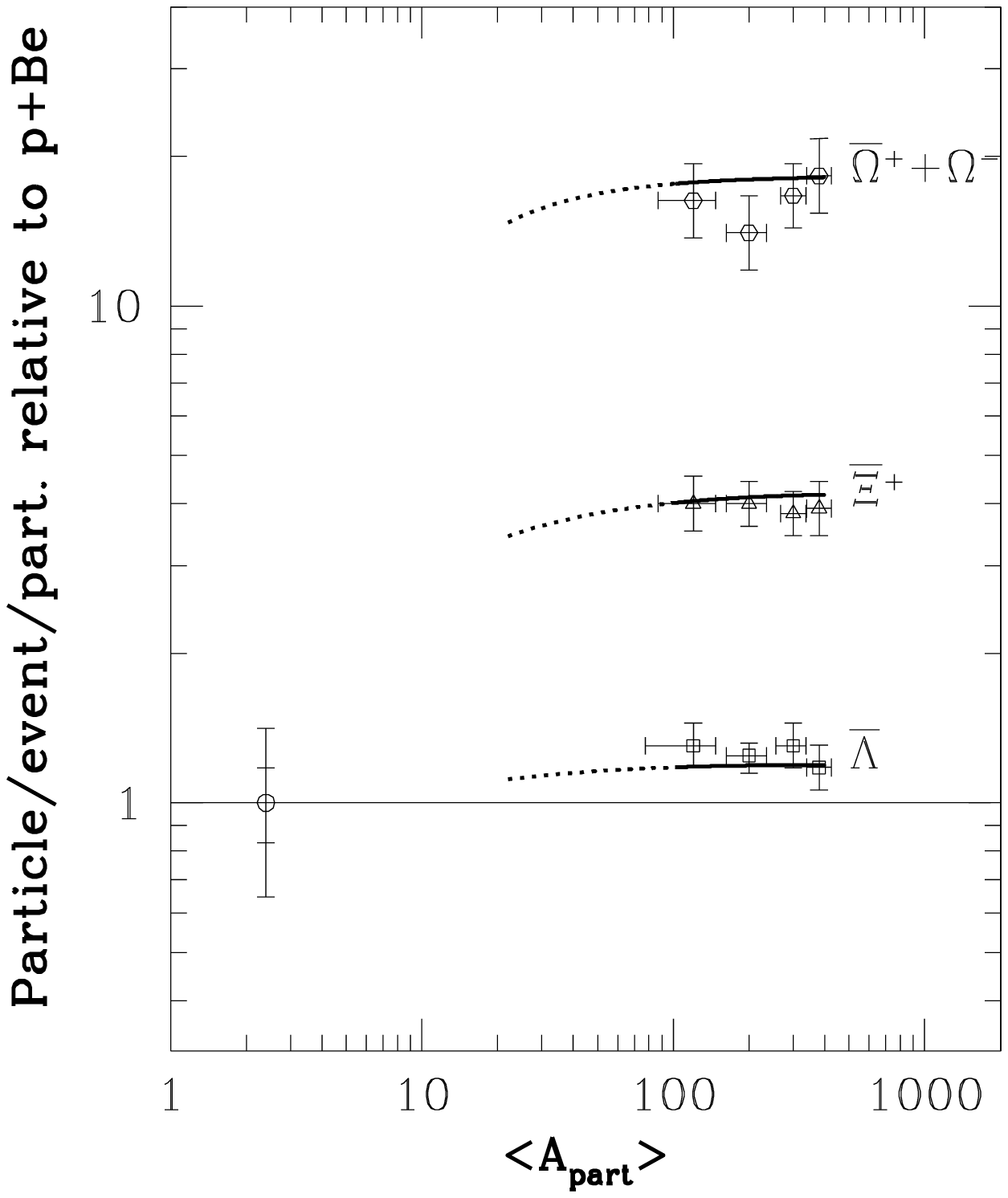}}
\vspace*{-.5cm}
\caption{ 
Particle yields/participant for strange baryons and antibaryons calculated in the canonical ensemble
with  $V_0\sim A_{part}/2$  
  normalized
to their values at $\bar V_0$. 
Also shown are the WA97 experimental
 data for yields  per
participant in Pb-Pb relative to p-Be \cite{Ant99}.
For description of  curves see text.
\protect\label{pen}}
\vspace*{-.2cm}
\end{figure}

In Fig.\ref{pen} we show the densities
 calculated in C
 ensemble for large $V_0$
normalized to its value at $\bar{V}_0$  in to p-Be system.
This corresponds to
 the experimental
comparison of yield/participant of Pb-Pb
to p-Be  system. The full lines in Fig.\ref{pen}
are calculated from Eq.(\ref{eq3})  by varying 
$V_0$  and keeping  $\mu =266$ MeV.
 The change 
of $V_0$ with centrality in Pb-Pb collisions is included relating
$V_0=4\pi R_0^3/3$ with $A_{part}$ through,
 $R_0\sim 1.2( A_{part}/2)^{1/3}$, which
determines approximately the number of projectile participants.

The actual path  from grand canonical to canonical
result for the  particle densities  would 
required an additional knowledge of $A_{part}$
dependence of chemical potential which is effectively  changing from
266 MeV in  Pb-Pb  to 150 MeV in   p-Be  system. Formulating,
in addition to strangeness, also 
baryon number conservation canonically,
the  path from GC to  C  result as a function of $V_0$
can be established \cite{sha}. In the simplified model, presented here,
where baryon number conservation was introduced effectively trough
 chemical potential we can only study the relative enhancement
of strange baryon densities from small p-Be  to large
Pb-Pb  system.

The result presented in Fig.\ref{pen} shows that the relative enhancement
of strange baryon and antibaryon yield/participant from  p-Be to   Pb-Pb, 
 found experimentally, can be well described
by the  effect of canonical suppression
and the change of $\mu$ from 266 MeV in  Pb-Pb
to 150 MeV in  p-Be collisions.
The enhancement appears in the model to be almost saturated
for $A_{part}\simeq 100$ which indicates that the system is close to its
GC limit. 
 The enhancement pattern for different strange 
baryons and antibaryons  densities is also understood within the  model
as a consequence of different dependence on $V_0$.
In the large system like Pb-Pb, and for large collision energy required
to maintain high $T$,
the density $n_s$ of particle  carrying strangeness $s$, 
 is $V_0$ independent. In the opposite limit, however, this
dependence is changed to $n_s\sim V_0^{s}$ which can be
verified from Eq.\,(\ref{eq3}). For small $x_i$ we have approximately:

\begin{equation}
n_{\pm s}  \simeq
Z_{\pm s} {{(S_{\pm1})^s}\over 
{{(S_{+1}}S_ {-1})^{s/2}}}~
      { { I_s(x_1)}\over {I_0(x_1)}}. 
\label{eq6}
\end{equation}
 Expanding the Bessel functions  $I_n(x)\sim x^n$ in Eq.\,(\ref{eq6})
 we see that 
 $n_s \sim V_0^s$.  

The variation  of $V_0$ with the number of projectile participants
and the assumption that $V\neq V_0$ is also justified by the data.
Indeed  the recent data of the WA97
collaboration show  that strange particle yield/$A_{part}$ is independent
on $A_{part}$ in p-A collisions. This is well reproduced by canonical
treatment since $V_0$ is the same for all A in p-A collisions.
If one could, however, take 
$V=V_0$ then, since $V\sim A_{part}$, 
 a strong variation with $A_{part}$ of strange
particle densities should be observed in p-A collisions when changing A. In addition this variation would
depends on the strangeness content of the particle which is not
experimentally observed in p-A.
 Thus from the point of view of the  considered canonical
 model, p-Be and p-Pb 
are the same system  with the  only difference  being  contained in
 slightly larger $\bar\mu$ 
 in p-Pb then p-Be. We found that the chemical potential
in p-Pb increases by 
about  10$\%$ with respect to p-Be. This explains similar excess
of particle yield/$A_{part}$ in  Pb-Pb relative to p-Be and p-Pb
collisions. 


Until now we have assumed that p-A  as well as Pb-Pb system 
can be described
by the statistical model in   chemical equilibrium.
For Pb-Pb collisions chemical equilibrium assumption has been found
to give a good agreement of the model with the experimental data \cite{Bra99}.
In p-A collisions, however, due to small time scale,  deviations
from strangeness chemical saturation cannot be excluded.
 In a previous analysis \cite{Bec96}
of particle production in ${\rm p}-{\rm p}$, ${\rm p}-\bar{\rm p}$ and $e^+-e^-$ collisions in terms of
 C ensemble  
strangeness under saturation was accounted for by introducing an
additional 
 parameter 
$\gamma$ which measures deviation of strangeness from full
 equilibrium \cite{Raf92}.
 Also in this approach $V_0$ was assumed to coincide with $V$.
 In the following we discuss 
how the possible out of chemical
 equilibrium together with  the $V=V_0\sim A_{part}$
 assumption 
will modify our results
and conclusions.

To introduce  non-equilibrium effects via $\gamma$ parameter is quite
straightforward \cite{Raf92,Let95}.
It amounts to the  replacement
 $S_{\pm s}\mapsto \gamma^sS_{\pm s}$
in Eq.\,(\ref{eq3}). First we observe that applying
the limit $x_i\mapsto 0$ in Eq.\,(\ref{eq6})
 the density $n_s$  of particle  with strangeness
$s$,  scales as $(\gamma^2V)^s$. For larger $x_i$ we found that, 
 in the parameter range relevant for p-A system,
 this dependence is   changed  approximately to 
 $n_s\sim (\gamma^{2.7}V)^s$.
 Thus, keeping $\gamma^{2.7}V=\bar{V}_0$, the value 
used in
equilibrium model, we can as well reproduce the experimental data shown in
 Fig.\ref{pen}
 in   non-equilibrium formulation.  
In this approach, however, in order to 
reproduce experimentally observed 
  linear dependence
of strange particle yields with $A_{part}$ in p-A collisions \cite{Ant99}, one
needs,  for each value of $A_{part}$, to adjust $\gamma$ such that
$\gamma^{2.7}A_{part}=const$. Obviously, this requires a
  decreasing $\gamma$
with $A_{part}$. In general, one would rather expect the opposite behavior. In order to ovoid, however, the possible ambiguity
with respect to strangeness undersaturation factor $\gamma$ 
one needs to perform an additional detailed analysis in proton
induced collisions for different targets and collision energies.
This analysis could possibly settle to what extend there is a need
for genuine global strangeness enhancement related with the change
in $\gamma$ between p-A and A-B collisions.
\section{Summary}
We have used the comprehensive set of the  recent results of the WA97
 collaboration  on the production of strange and multistrange hadrons
 in p-A and Pb-Pb collisions  in order to check
the validity of statistical model description of 
 strangeness enhancement centrality dependence.  
The statistical  model was verified to give a good 
agreement  with all experimental data in A-A collisions at SPS energies. 
 From the point of view
of statistical mechanics the application of the  model for smaller system
like p-A requires obvious modifications. In the large system  strangeness
is conserved on the average, thus it is controlled by strange chemical potential
in the grand canonical ensemble.
 In the small system, however,
 strangeness conservation laws must strictly
 be obeyed  in each
single collision which requires canonical ensemble.
We have shown that the required passage from grand canonical
 to canonical ensemble
reproduces
 the basic features of strangeness suppression from Pb-Pb to p-Be or p-Pb collisions reported by WA97 collaboration. The observed asymmetry
in the enhancement of strange   baryons and antibaryons requires additional
assumption on $A_{part}$ dependence of baryon chemical potential $\mu$.
We have shown that  with the particular choice of  $\mu$ asymmetry in
p-A and Pb-Pb systems the  model  provides   good quantitative
 description of the data.

Deviation from chemical equilibrium has been included in our
 discussion only in p-~A system. Recently an interesting  non-equilibrium model 
for the description of particle production in A-B collisions has
been proposed \cite{Raf00}. The interpretation 
of centrality dependence of strangeness enhancement
in a consistent non-equilibrium picture would require additional
analysis not presented in this paper.


\subsection*{Acknowledgment}
\hspace*{\parindent}
We  acknowledge  stimulating
discussions with Rolf Hagedorn and Ulrich Heinz.
One of us (K.R) also acknowleges interesting comments
by E. Suhonen and the partial support
of LPTHE,  Universit\'e Paris 6 and 7, Unit\'e Mixte de Recherche
du CNRS, UMR 7589, 
and of Committee for Scientific Research (KBN-2P03B 03018).



\begin{thebibliography}{99}
\bibitem{qm99ski}
          { {\it Quark Matter '99}},
          Proc. of the XIV'th Int. Conf. on
          Ultra-Relativistic Nucleus-Nucleus Collisions,
          Torino, Italy, 
  Nucl. Phys. {\bf A661} (1999).
 
\bibitem{Ant99} F. Antinori, $et$ $al.$, WA97 collaboration, Nucl. Phys. {\bf A661}, 130c (1999).

\bibitem{Lie99}R. Lietava, $et$ $al.$, WA97 collaboration,
J. Phys. {\bf B25}, 181 (1999).

\bibitem{And298} E.Andersen, $et$ $al.$, WA97 collaboration,
Phys. Lett. {\bf B449}, 401 (1999).

\bibitem{Sik99} F. Sikler, $et$ $al.$, NA49 collaboration,
Nucl. Phys. {\bf A661}, 45c (1999).

\bibitem{Bra99} P. Braun-Munzinger, I. Heppe, and
J. Stachel,
Phys. Lett. {\bf B465}, 15 (1999).
\bibitem{Bra95} P.  Braun-Munzinger,  J. Stachel, J. P. Wessels,
and N. Xu, Phys. Lett. {\bf B344}, 43 (1995);
 Phys. Lett. {\bf B365}, 1 (1996).

\bibitem{CLK} J.  Cleymans, and K. Redlich,
Phys. Rev. Lett. {\bf 81}, 5284 (1998) {\it and references therein.} 

\bibitem{he} U. Heinz,  Nucl. Phys. {\bf A661}, 349 (1999);
   R. Stock, Phys. Lett. {\bf B456}, 277 (1999).

\bibitem{Let95} J. Letessier, A. Tounsi, U. Heinz, J. Sollfrank, and
J. Rafelski, 
Phys. Rev. {\bf D51}, 3408 (1995).

\bibitem{Cle99} J.  Cleymans, and K. Redlich,
Phys. Rev. {\bf C60}, 054908 (1999); 
J. Cleymans, H. Oeschler, and K. Redlich,
Phys. Rev. {\bf C59}, 1663 (1995); nucl-th/0004025.

\bibitem{Bec96}
F. Becattini,  Z. Phys. {\bf C69}, 485 (1996);
F. Becattini, and U. Heinz, Z. Phys. {\bf C76}, 269 (1997).
 
\bibitem{Hag71}R. Hagedorn, CERN yellow report 71-12, 101 (1971).

\bibitem{Raf80} J. Rafelski, and M. Danos,
 Phys. Lett. {\bf B97}, 279 (1980).

\bibitem{Hag85}R. Hagedorn, and K. Redlich
, Z. Phys. {\bf A27}, 541 (1985). 

\bibitem{Cle91}J. Cleymans, K. Redlich, and E. Suhonen,
, Z. Phys. {\bf C51}, 137 (1991).
 
\bibitem{Cle96}J. Cleymans, and A. Muronga,
Phys. Lett. {\bf B388}, 401 (1996); 
J. Cleymans, M. Marais and E. Suhonen,
Phys. Rev. {\bf C58}, 2747 (1997);
J. Cleymans, D. Elliott, A. Ker\"anen, and E. Suhonen,
Phys. Rev. {\bf C57}, 3319 (1998).

\bibitem{Sol98}
J. Sollfrank, F. Becattini, K. Redlich, and  H. Satz,  Nucl. Phys. {\bf A638}, 399c (1998).

\bibitem{clT} F. Becattini,  M. Gazdzicki, and J. Sollfrank,
 Eur. Phys. {\bf JC5} 143 (1998) 
\bibitem{sha} S. Hamieh, K. Redlich, and A. Tounsi {\it to appear}.

\bibitem{Raf92} J. Rafelski, Phys. Lett. {\bf B262}, 333 (1991).

\bibitem{Raf00} J. Letessier, and J. Rafelski, nucl-th/0003014,
Int. J. of Mod. Phys.  {\bf E} (2000) {\it to appear.} 
 
\end{thebibliography}
\end{document}